\documentclass[onecolumn,showpacs]{revtex4}

\usepackage{graphicx}%
\usepackage{dcolumn}
\usepackage{amsmath}

\makeatletter
\def\btt#1{\texttt{\@backslashchar#1}}%
\DeclareRobustCommand\bblash{\btt{\@backslashchar}}%
\makeatother

\begin{document}


\title{Enhancing Curiosity Using Interactive Simulations Combined with Real-Time Formative Assessment Facilitated by Open-Format Questions on Tablet Computers}

\author{F.V. Kowalski and S.E. Kowalski}
\affiliation{Physics Department, Colorado
School of Mines, Golden CO. 80401 U.S.A.}

\begin{abstract}
Students' curiosity often seems nearly nonexistent in a lecture setting; we discuss a variety of possible reasons for this, but it is the instructor who typically poses questions while only a few students, usually the better ones, respond.

As we have developed and implemented the use of InkSurvey to collect real-time formative assessment, we have discovered that it can serve in an unanticipated role: to promote curiosity in engineering physics undergraduates. Curiosity often motivates creative, innovative people. To encourage such curiosity, we solicit questions submitted real-time via InkSurvey and pen-enabled mobile devices (Tablet PCs) in response to interactive simulations (applets) run either before or in class. This provides students with practice in asking questions, increases metacognition, and serves as a rich springboard from which to introduce content and/or address misconceptions.

We describe the procedure for measuring curiosity and results from applying this method in a junior level electromagnetics engineering physics course.  We conclude that students are indeed more curious than they appear in class, and students participate even without extrinsic motivation. This method of enhancing curiosity using interactive simulations coupled with real-time formative assessment in response to open-format questions could be implemented in a wide variety of science and engineering courses as well as elsewhere.

\end{abstract}


\maketitle

\section{Introduction}
\label{sec:intro}

Most people would probably agree that good engineers and scientists are curious.  An emerging operational definition of curiosity is ``the threshold of desired uncertainty in the environment that leads to exploratory behavior" \cite{jirout}.  In spite of recent advances in cognitive science and learning theory \cite{bransford}, many engineering educators find themselves teaching in a lecture setting, where students' curiosity and intellectual exploration is rarely displayed. This lack of evidence of students having crossed the threshold of curiosity might have several possible contributing factors;  our experiences suggest these factors may include:

(a) Students may be embarrassed to ask questions (either in the presence of their peers or in a one-on-one setting with the instructor during office hours);

(b) Students may not want to make a genuine effort to understand the material by asking probing questions;

(c) Students' past experiences may have trained them to expect to absorb content rather than generate questions;

(d) Students may feel so lost and confused that no question comes to mind;

(e) Students may feel confident that they understand everything, and therefore have no need to ask questions; and

(f) Students may be accustomed to not asking questions, as this is not a skill generally nurtured or valued in the educational system.

However, curiosity often motivates creative, innovative scientists and engineers (and others) \cite{burleson,koestler,simon}. It is also seen as an important factor in student learning \cite{day,mcnay}. Practical methods for encouraging curiosity have been discussed \cite{day,tomkins,vidler,amone}. Lowenstein \cite{loewenstein} points out that ``curiosity is influenced by cognitive variables such as the state of one's knowledge structures but may, in turn, be one of the most important motives for encouraging their formation in the first place. Positioned at the junction of motivation and cognition, the investigation of curiosity has the potential to bridge the historical gulf between the two paradigms."

This paper describes results when students in a junior-year undergraduate electromagnetics class were asked to report their questions as they explored interactive simulations related to course content.  The goal was to motivate learning by enhancing student curiosity.  To better understand the role curiosity plays in engineering education, a categorization scheme for questions was devised and applied as a simple  measurement tool.

\section{NURTURING AND MEASURING CURIOSITY}
\label{sec:nurturing}

Efforts to improve student motivation have been linked with curiosity. In particular, the ARCS (Attention, Relevance, Confidence, and Satisfaction) model of motivation provides insight on how to nurture curiosity \cite{amone}. Some instruments to measure motivation have curiosity subscales \cite{day2}, reflecting the contribution curiosity may make to motivation. However, curiosity is not a simple attribute to measure. Loewenstein provides a comprehensive review of instruments which use teacher evaluations, peer evaluations, and self-evaluations to measure curiosity, while he questions the validity of some of these instruments \cite{loewenstein}.

Trying to understand the underlying cause of creativity has led to theories focused around drive (curiosity results from biological/psychological drives similar to thirst, hunger, or sex), incongruity (curiosity results when an observation violates a model or heuristic rule of how the world works), competence motive (curiosity results from feelings of a lack of competence), and a gap in information  (curiosity results from a discrepancy between what one perceives and what one expected to perceive) \cite{loewenstein,litman}.

Rather than focus on a comprehensive study of the underlying causes of curiosity, we attempt to extract information about a student's curiosity only in terms of the questions the student asks about an observation, problem, simulation, or object presented in a STEM context.

In theories of curiosity described in the literature, the types of questions students might ask are not categorized. For example, in an incongruity-based theory, such as that developed independently by Hebb \cite{hebb}, Hunt \cite{hunt}, and Piaget \cite{piaget}, any question reflecting curiosity would, by construction of the theory, be defined as incongruous.

None of these theories about curiosity has emerged as the most widely accepted, but perhaps there are insights to be gained from all of them. An example, in the context of Newton's Laws, illustrates that student questions do not always conform with a single theory.  A student who does not believe the rule that an object moves in uniform motion until an external force acts would ask a question such as ``What object moves without slowing down?"  This aligns with the incongruous theory of curiosity (above), since it indicates that uniform motion is incongruous with the student's model of how the world works.  Another student question might be ``How do I calculate the external force on a car coasting?" illustrating curiosity about how to apply the rule rather than about a violation of the rule.  For convenience, this could be described as a question reflecting congruous curiosity.

Based on this theoretical framework, the question categories used in this study to assess curiosity are as follows. Although this list of categories may not be exhaustive for consideration of curiosity in all disciplines or even all topics within engineering disciplines, they are adequate for this study.\\*
\\*\underline{INCONGRUOUS}: violating a model or heuristic rule of how the world functions.

Example: Doesn't this violate conservation of energy?\\*
\\*\underline{CONGRUOUS}: understanding of or gathering information about how a model or rule is applied (typically one just presented in class).

Examples: How do I apply this in a different situation? How do I calculate the effect shown in the simulation? What examples are found in the real world?\\*
\\*\underline{MODIFYING}: probing what happens when the assumptions, parts, application, or parameters of the model or rule are changed.

Example: What happens if the temperature is not assumed to be constant?\\*
\\*\underline{GENERALIZING/ANALOGY}: comparing one model with another.

Example: When an earthquake occurs, do the plates slip when pressure generates melting at an interface, like a skate on ice?\\*
\\*\underline{CAUSAL/CREATIVE}: attempt to generate a new model, improve on an existing one, or search for novel patterns.

Example: How does this simulation of classical physics change if quantum mechanics is applied?\\*
\\*\underline{INFORMATIONAL}: finding information simply for its intrinsic interest or for diagnostic purposes.

Example: How does the platypus relate to its environment?\\*

When the student responses are collected as students explore interactive simulations, we categorize the question responses and discuss in class these classifications with the students. Using this procedure, along with having the students become aware of questions asked by both peers and the instructor, we hope to nurture the transformation of specific state curiosity (a condition that can be manipulated) into the more general trait curiosity (a personality feature) in STEM fields, as suggested by Loewenstein \cite{loewenstein}.

\section{Description of Teaching Model}
\label{sec:model}

Students used the web-based software InkSurvey \cite{ticc} to compose and submit their responses to questions about free, interactive simulations they accessed online. There are a variety of tools that could be used for this (DyKnow, Classroom/Ubiquitous Presenter, etc.); we chose InkSurvey because it is free and is robust in classes exceeding sixty students. The preparation, collection, and receiving of responses was facilitated by pen-enabled mobile computing devices (in this case tablet computers) in the hands of both instructor and students. Other pen-enabled mobile computing devices (iPads, smart phones, etc.) could also be used in a similar manner. However, without facilitation by some form of technology, it is difficult for the instructor to collect and respond to meaningful formative assessments quickly.

The real-time assessment gathered with InkSurvey is particularly authentic since it is seamlessly integrated with the activity \cite{young}, in this case not only temporally, but physically as well, with a single tool used for both exploring simulations and using InkSurvey. An additional advantage of this coupling is that students can engage in further manipulation of the simulations as they construct their responses, allowing them to explore facets they may have initially ignored. A single student is able to submit multiple responses to a question posed by the instructor.  Even though student identity is concealed from their peers (see I.a), the instructor  could access this, thus allowing us to track how many and which responses were submitted by a particular individual.

The student responses provide a rich springboard from which to introduce content and/or address misconceptions. Having the students document their questions and understanding in this manner facilitates both student motivation to understand and metacognition.

In this method of enhancing curiosity, comments and questions were solicited using InkSurvey, with open-format questions such as:

(1) What did you learn from or observe in the simulation?

(2) What does this simulation illustrate?

(3) What are you curious about after running the simulation?

The students enrolled in the class were engineering physics second semester juniors having three semesters each of physics and calculus, along with courses in differential equations, linear algebra, intermediate mechanics, thermodynamics, analog and digital electronics, and a summer session in vacuum systems, optics, machine shop, computer interfacing, and electronics.

\section{Results}
\label{sec:results}

Student responses from four interactive simulations, along with categorizations and instructor comments, are presented next.  To make trends more visible, similar responses are reported together, but unique responses are also reported to show the level and breadth of curiosity within the class. However, student submissions do not represent all categories for every simulation.  Every student present at the time of each simulation submitted at least one response.

\subsection{Induced charge simulation}

Students ran the simulation on electrostatic induction using the spherical and unsymmetrical conductors found here:\\*
\url{http://www.cegep-ste-foy.qc.ca/freesite/index.php?id=1826}\\*

Question 3 above (``What are you curious about after running the simulation?") elicited 68 questions from the 40 students present.  These questions fall into five of the possible six categories.\\*
\\*\underline{INCONGRUOUS}.    Two responses are in this category:
\\*``Does this actually happen as shown?"
\\*``Why don't the charges move after the conductors stop moving?"\\*
\\*\underline{CONGRUOUS}. Twenty-four student questions (35\% of total) involve how the charge distributed over the conductors, as shown in the simulation.  Sixteen student responses (24\% of total) involve how much charge was shared between conductors. Another eight student questions (12\% of total) are related to how the charge distribution, electric fields, or forces could be calculated.  Unique responses in the CONGRUOUS category are:
\\*``How do the charges of opposite sign eliminate each other?"
\\*``What is the charge transfer a function of?"\\*
\\*\underline{MODIFYING}. Six of the student questions (9\% of total) involve what happens if the "tear shaped" conductor is turned around or two such conductors are used.  Additionally, there are three unique responses in the MODIFYING category:
\\*``How does the shape of the conductor affect the charge distribution?"
\\*``How would insulators effect the simulation?"
\\*``What happens if the conductors are brought together a second time?"\\*
\\*\underline{CAUSAL/CREATIVE}. There are no questions in this category that were submitted multiple times, but these unique questions were received:
\\*``How much time does it really take for the charges to distribute themselves over the conductor?"
\\*``How does this simulation differ if quantum mechanics is applied?"
\\*``Can electrons jump between the conductors or do the conductors need to touch?"\\*
\\*\underline{INFORMATIONAL}.  Four responses (7\% of total) are questions about clarification, such as ``Is there charge missing on the right conductor initially?" Another student questioned, ``Where do the charges come from?"

Comments: While the majority of questions focus on what happens in the simulation, some students were curious about how to calculate what happens. A major objective of the course is for students to learn how to perform such calculations. The simulation thus provided motivation for the mathematical content of the course.\\*

The students were concurrently taking a quantum physics class. A few were curious about how the content in both classes related. The quantum and classical models of charged conductors interacting are not often discussed at the junior level. By coupling the two topics, the breadth of understanding increases, thereby enhancing capability to both construct models and understand the limitations of the models. This simulation can also be used as a springboard to discuss quantum tunneling.

Some students were also curious about how quickly the charges distributed over the conductors. While this is not a question which can be adequately addressed using the electrostatic model presented, the instructor can use this as a catalyst to question the limitations of the electrostatic model, and launch a discussion of what governs the dynamics of this process and how one would construct a more realistic model.

\subsection{ Moving charge simulation}

Students ran the simulation on electric fields from a moving charge found here:\\*
\url{http://www.cco.caltech.edu/~phys1/java/phys1/MovingCharge/MovingCharge.html}.\\*

Subsequently asking the students, ``What are you curious about after running the simulation?" elicited 53 questions from the 31 students present. These student questions fall into five of the possible six categories.\\*
\\*\underline{INCONGRUOUS}.   Six responses (11\% of total) deal with how the magnetic field (not shown in the simulation) is generated. Four student responses (8\% of total) question how it is that field lines can cross (shown when running the saw-tooth motion option). Unique responses are:
\\*``When the charge is moving fast why doesn't E continue in front of the charge?"
\\*``Why is the charge able to outrun the E field when traveling at the speed of light?"
\\*``Why are there kinks in the field lines when the charge accelerates?"
\\*``Won't the waves always move faster than the particle?"
\\*``Why does acceleration cause ripples?"
\\*``What makes the electric field E behave like a wave?"
\\*``Why does the electric field seem to get stronger towards the front as the charge goes faster?"
\\*``Why is there an asymmetry in the field between the front and back of the moving charge?"
\\*``What happens when the particle crosses a bending E field?"
\\*``How can the field lines cross over each other when selecting the saw-tooth motion?"
\\*``What is the mechanism behind the field distortions when the particle speed changes?"\\*
\\*\underline{CONGRUOUS}. Seven responses (13\% of total) deal with "compression" of the field lines when the charge moves at high constant speed. Unique questions generated by the students are:
\\*``How is energy shown?"
\\*``How can you calculate the kinks in the field?"
\\*``Do the distorted field lines create magnetic fields?"
\\*``How can we calculate the fields given the particle motion?"
\\*``Is the superposition principle used to calculate the effect?"
\\*``How does the field propagate when the charge undergoes simple harmonic motion?"
\\*``Is there any change in the amplitude or frequency of the electromagnetic waves produced as they propagate away from the source?"
\\*``How do we account for the force based on the speed of the moving wave?"\\*
\\*\underline{MODIFYING}. Six student questions (11\% of total) are concerned about what happened when the charge moved at or near the speed of light.  Additional unique student questions received are:
\\*``How does this simulation change in a gravitational field?"  \\*``What happens when two particles accelerate?"
\\*``What are the relativistic effects?"
\\*``How does the magnetic field affect motion of the particle?"
\\*``How does the magnetic field arise?"
\\*``Are there kinks in the magnetic field when the charge accelerates?"
\\*``How does this moving particle interact with other particles during acceleration?"\\*
\\*\underline{GENERALIZING/ANALOGY}: There are three unique questions that fall in this category:
\\*``When the charge moves faster than the speed of light, is this similar to a bow wave when an aircraft moves faster than sound?"
\\*``How does this simulation apply to light?"
\\*``Are there shock waves generated?"\\*
\\*\underline{CAUSAL/CREATIVE}. A single student question is classified in this category:
\\*``Do multiple charges create an interference pattern?"

Comments: Some students noticed that under certain conditions the simulation generates crossing field lines. They correctly suspected that this cannot happen physically. Bringing this to the attention of the class facilitates a discussion of critical thinking skills about what methods can be used to verify models (or how the model is simplified to generate the simulation).

Questions about the interference pattern illustrate the ability to make connections between different physical phenomena. The analogy with shockwaves can be a springboard to generate more questions about what is similar or different between the two situations. The connection between acoustics and electromagnetic waves can be discussed in terms of the similarity between the partial differential equations used in each model.

\subsection{Inductance calculator}

Students ran the inductance calculator found here:\\*

\url{http://emclab.mst.edu/inductance/index.html}.\\*

The question ``What did you learn in using the calculator?" elicited 35 comments from the 35 students present; 100\% of the responses submitted fall into the CONGRUOUS category. Seventy-eight percent (78\%) mention the relationship between geometry and inductance while seven percent (7\%) relate this to the induced electromotive force changing as a function of geometry.  A typical comment is ``Different shapes make the calculations way harder. It is cool that we know how to calculate this." Curiosity is apparent in the following two comments: ``I didn't realize that for the two loop configuration that the inductance would increase with the separation distance," and ``The $N^{2}$ relationship was unexpected. However when we look closer at the math I understand."
Comments: Somewhat surprisingly, students found this calculator interesting. They demonstrated curiosity in exploring how inductance varied in different geometries. The limited number of geometries available provided an opportunity to discuss the scarcity of closed form solutions for inductance.

\subsection{Quantum harmonic oscillator simulation}

Students ran the simulation found here:\\*

\url{http://web.ift.uib.no/AMOS/MOV/HO/}.\\*

The prompt ``What are you curious about after running the simulation?" elicited 32 questions from the 31 students present.\\*
\\*\underline{INCONGRUOUS}.  Ten responses (30\% of total) deal with ``Why are we studying this problem in a electromagnetism class?"  Additional unique student questions that fall into this category include:
\\*``Why don't we see evidence of this non-classical behavior on a macroscopic scale?"
\\*``Can we make such a state?"
\\*``Why does the oscillator look more classical with a greater number of energy levels?"\\*
\\*\underline{CONGRUOUS}.   Unique student questions in this category are:
\\*``Why are coherent states so important?"
\\*``What gives them a stable shape?"
\\*``What does it mean when the shape of the wavefunction splits and recombines?"
\\*``Does this correspond to a transition to a higher energy level?"
\\*``Can we use the superposition principle?"
\\*``Why did the wave always move to the right?"
\\*``How accurate is the simulation?"
\\*``What are the limitations of the simulation?"
\\*``What is represented by non-coherent states?"
\\*``Why is the Glauber state important?"
\\*``What impact do energy levels not excited have on the state?"
\\*``How do you produce just even or odd wavefunctions?"
\\*``Why do so many amplitudes have to exist to get a standard wavepacket?"
\\*``Is this an example of multiple electrons or single electrons?"\\*
\\*\underline{MODIFYING}. Two student questions belong to this category:
\\*``What simplifications went into making the simulation?"
\\*``How would this behave if the oscillator were damped of experienced an external force?"\\*
\\*\underline{GENERALIZING/ANALOGY}:  There are two student questions in this category:
\\*``Is this how lasers work?"
\\*``Are there any real world electromagnetic examples of these states?"

Comments: Students explored this simulation after having studied the separation of variables solution to Laplace's equation. The harmonic oscillator simulation is a quantum example of separation of variables applied to the Schrödinger equation. This topic was being studied by most of the members of the electromagnetics class in a concurrent quantum physics class. Some students did not see the mathematical connection between the electrostatic and quantum problems. They
therefore asked why quantum content was being covered in an electromagnetics class.\\*

Many questions were focused on making connections between the simulation and real world applications. Answering these questions allowed the instructor to further stimulate curiosity and make connections to other physical phenomena. How a laser generates ultra short optical pulses in a manner similar to that demonstrated in the simulation is such an example.
Table I summarizes the results of categorizing the questions submitted as student responses for each of the four simulations described.

\section{Discussion}
\label{sec:discussion}

The use of pen-enabled mobile technology to collect student responses allows the instructor to receive student input in real-time and use it to guide students as they construct deeper understanding. This novel method of assessing and nurturing curiosity when exploring interactive simulations could be used effectively in many settings. Further explorations in contexts beyond those of using computer simulations are warranted by these results.

The results from classifying student responses, shown in Table I, are heavily predominated by curiosity associated with the students constructing an explanation of the simulation based on their understanding of the model presented in class (CONGRUOUS). The next largest contribution is associated with aspects of the simulation which do not match the student's expectations (INCONGRUOUS). Few students show curiosity about generalizing or modifying the model.

It is expected that students struggling to understand a model would have more congruous questions, focused on the details of calculating the consequences of the model, while those with a more mature understanding of the model would be curious about applications, analogies, modifying, or generating new models. Also, we suggest that the concepts and content illustrated by various simulations could influence the particular categories of curiosity they stimulate.

In this study, we have not addressed the issue of the ``quality" of the questions and have considered all student responses to be equal in the level of curiosity and insight they reflect.

\begin{table}[ht]
\caption{SUMMARY  OF  CATEGORIZATION  OF  STUDENT   QUESTIONS SUBMITTED,  FOR  EACH  OF  4  SIMULATIONS}
\centering
\begin{tabular}{c c c c c}
\hline\hline
Question & Sim A & Sim B & Sim C & Sim D \\[0.5ex]
\hline
Incongruous & 3\% & 40\% & 0\% & 40\% \\
Congruous & 74\% & 28\% & 100\% & 44\%\\
Modifying & 12\% & 25\% & 0\% & 6\%\\
Generalizing/Analogy & 0\% & 5\% & 0\% & 6\%\\
Causal/Creative & 4\% & 2\% & 0\% & 0\%\\
Informational & 7\% & 0\% & 0\% & 0\%\\ [1ex]
\hline
\end{tabular}
\label{table:nonlin}
\end{table}

The questions generated in these exercises are remarkable for their quantity, quality, and diversity, especially in light of the fact that students were given no external incentives for their responses. This is consistent with curiosity being an intrinsic motivational factor. Furthermore, none of the exams or homework exercises involved nurturing or assessing curiosity. From the students' questions and comments, it was apparent that they were genuinely interested in engaging with the material at a level not apparent prior to the real-time formative assessments.

The responses revealed an aspect of student thinking not often measured or even exposed in a lecture-based classroom. Many provided springboards for making connections between physical phenomena, discussing critical thinking skills, and enhancing curiosity.

Disadvantages of categorizing the types of curiosity include the non-uniqueness of the sorting categories and the potential confusion due to the redefinition of terms by which to sort.
One advantage of categorizing the questions is that students become aware of the many ways that they might be curious. This awareness can lead to a process which stimulates their curiosity. Consider a heuristic rule from a statics class that structures built in triangles are less likely to collapse. When asked what they are curious about, the students might apply the ``Modifying" category to ask how the skull of a reptile exhibits this rule. A student's curiosity could expand from state to trait as they are exposed to multiple examples in class.  Facilitated by technology such as InkSurvey, this process can be demonstrated by both the student's peers and the instructor. Hopefully, if this process is implemented often enough, students would become fluent in generating questions reflecting all categories of curiosity.

Additionally, the development of the method presented here to measure curiosity, if both valid and reliable, may lead to improved curricula. Categorization allows different levels and types of curiosity to be studied.

Other advantages of this teaching model are: (1) it reinforces the congruent category of curiosity, which aligns directly with the content STEM students are generally considered to be in class to learn about (they should be curious about how to calculate the consequences of a model they are studying, for example), (2) it encourages students to communicate what they think is incongruent so instruction can be modified to address such issues, and (3) it affords students practice in metacognition as they construct questions about which they are curious.

\section{Conclusions}
\label{sec:conclusions}

When students were provided with an opportunity to submit questions they generated while exploring interactive computer simulations, it was revealed that they are indeed more curious than they appear in a lecture-based class. They were willing to actively engage with the material they were studying, even without extrinsic motivation. Furthermore, the questions generated by individuals in the class provided relevant and worthwhile springboards for subsequent class discussions involving higher level thinking skills, as well as stimulating student curiosity to further explore the simulations.  The process and classification scheme described here may prove valuable in nurturing and measuring curiosity.

\end{document}